\documentclass[11pt,a4paper]{article}
\usepackage{CJK}
\usepackage{jstyle}
\usepackage{amsmath,amsfonts,amssymb}
\usepackage{simplewick}
\usepackage{graphicx}
\usepackage[dvipsnames]{xcolor}
\usepackage{float}
\usepackage{caption}
\usepackage{subcaption}
\usepackage{booktabs}
\usepackage{multirow}
\usepackage{tikz}
\usetikzlibrary{calc,matrix,decorations.pathmorphing,decorations.markings,arrows,positioning,intersections,mindmap,backgrounds,patterns,arrows.meta,chains}
\usepackage{cleveref}
\crefname{subsection}{subsection}{subsections}
\Crefname{subsection}{Subsection}{Subsections}

\DeclareGraphicsExtensions{.pdf,.png,.jpg,.mps}
\newcommand{\tr}{\operatorname{tr}}

\makeatletter
\DeclareFontFamily{OMX}{MnSymbolE}{}
\DeclareFontShape{OMX}{MnSymbolE}{m}{n}{
    <-6>  MnSymbolE5
   <6-7>  MnSymbolE6
   <7-8>  MnSymbolE7
   <8-9>  MnSymbolE8
   <9-10> MnSymbolE9
  <10-12> MnSymbolE10
  <12->   MnSymbolE12
}{}
\DeclareSymbolFont{MnLargeSymbols}{OMX}{MnSymbolE}{m}{n}

\let\llangle\@undefined
\let\rrangle\@undefined
\DeclareMathDelimiter{\llangle}{\mathopen}{MnLargeSymbols}{'164}{MnLargeSymbols}{'164}
\DeclareMathDelimiter{\rrangle}{\mathclose}{MnLargeSymbols}{'171}{MnLargeSymbols}{'171}
\makeatother

\title{Negative Rank Relations for  Giant Graviton Correlators}
\author[]{Qi Chen,}
\author[]{Zhongjie Huang,}
\author[]{Xinan Zhou}
\affiliation[]{Kavli Institute for Theoretical Sciences, University of Chinese Academy of Sciences, Beijing 100190, China}
\emailAdd{}{chenqi25@ucas.ac.cn,huangzhongjie@ucas.ac.cn,xinan.zhou@ucas.ac.cn}
\abstract{We study correlation functions of giant and dual giant gravitons in $\mathcal N=4$ super Yang--Mills theory. Although the two objects correspond to rather different D3 brane configurations in $AdS_5\times S^5$, their gauge theory operators are Schur polynomials associated with transposed Young diagrams. We show that this transposition is accompanied by a formal rank reflection $N\to -N$ at the level of correlation functions. For both $U(N)$ and $SU(N)$, the relation is exact for free correlators at finite $N$, and, using Lagrangian insertion, extends to separated point weak coupling integrands at every loop order. More generally, it applies to any number of Schur polynomial external operators, provided all Young diagrams are transposed. We illustrate the relation in several examples and also discuss evidence for its extension beyond perturbative integrands.
}

\begin{document}

\begin{CJK*}{UTF8}{}
    \CJKfamily{gbsn}
    \maketitle
\end{CJK*}

\tableofcontents

\newpage

\section{Introduction and summary}

Giant and dual giant gravitons are two rather different D3 brane realizations of heavy $\frac{1}{2}$-BPS states in $AdS_5\times S^5$. A giant graviton wraps a three sphere inside $S^5$, while a dual giant wraps one inside $AdS_5$ \cite{McGreevy:2000cw,Grisaru:2000zn,Hashimoto:2000zp}. In the gauge theory, however, the corresponding operators are closely related Schur polynomials \cite{Balasubramanian:2001nh,Corley:2001zk}: a giant of charge $M$ is associated with the one-column Young diagram $[1^M]$, whereas a dual giant is associated with the transposed one-row diagram $[M]$. Equivalently, their generating functions are a determinant and an inverse determinant. This contrast suggests a simple question: to what extent does Young diagram transposition relate the dynamical observables in the two heavy sectors?

We use heavy-heavy-light-light (HHLL) correlators as the main setting in which to address this question. The two heavy operators have charge $M=O(N)$, while the two light operators are protected single-particle operators of charges $p$ and $q$. Correlators of light $\frac{1}{2}$-BPS operators in $\mathcal N=4$ SYM have been studied extensively at weak coupling (see, e.g., the review \cite{Heslop:2022xgp}), at strong coupling \cite{Freedman:1998tz,DHoker:1999pj,Arutyunov:2000py,Rastelli:2016nze,Rastelli:2017udc} (see also \cite{Zhou:2026xtp} for a recent review), and at finite coupling \cite{Basso:2015zoa,Fleury:2016ykk,Coronado:2018cxj,Belitsky:2020qrm,Binder:2019jwn,Chester:2020dja,Dorigoni:2021bvj,Dorigoni:2021guq}. A complementary line of work has studied correlators involving heavy operators, and in particular those describing giant gravitons. These correlators have been investigated using representation theory, integrability, localization and bootstrap methods \cite{Balasubramanian:2001nh,Corley:2001zk,Berenstein:2003ah,Bissi:2011dc,Jiang:2019xdz,Jiang:2019zig,Chen:2019gsb,Vescovi:2021fjf,Yang:2021kot,Berenstein:2022srd,Holguin:2022drf,Lin:2022gbu,Jiang:2023uut,Brown:2024yvt,Chen:2025yxg,Brown:2025huy,Wu:2025ott,Chen:2026ium,Brown:2026dhy,He:2026ios,Chen:2026fnf}. One important distinction of the giant regime is that each heavy operator contains $O(N)$ fields, so nonplanar color contractions need not be suppressed in the usual way.

At the operator level, Young diagram transposition directly relates giant and dual giant operators. However, it does not by itself imply a relation between correlators. Already in the free theory, Wick contractions close additional color index loops and generate powers of $N$. At higher orders there are further color contractions from interaction vertices. Remarkably, we find that Young diagram transposition is compatible with Wick contractions under the following reflection
\begin{equation}
  \tr \longrightarrow -\tr,\qquad  N\longrightarrow -N\,.
\end{equation}
We refer to this as the {\it negative rank relation}. For the HHLL correlators we find
\begin{align}
  \widetilde G^{(0)}(N;M,p,q) &= (-1)^{M+(p+q)/2} G^{(0)}(-N;M,p,q)\,,  \\
  \widetilde{\mathcal I}^{(L)}(N;M,p,q) &= (-1)^{M+(p+q)/2+L} \mathcal I^{(L)}(-N;M,p,q)\,,
\end{align}
where functions with and without tildes are associated with dual and sphere giant gravitons respectively. The first relation is about correlators in the free theory and is exact at finite $N$. The second extends the result to integrands at every loop order through Lagrangian insertion, provided all external and insertion points are mutually separated. Both statements hold for $U(N)$ and $SU(N)$ gauge groups. In fact, the HHLL relation is a special case of a more general statement applying to any number of Schur-polynomial external insertions, with every Young diagram transposed. Throughout, $N\to -N$ is understood as a formal continuation of the color factors.

The underlying color algebra mechanism is closely related to the classical negative dimensionality theorem \cite{Cvitanovic:1982bq,Elvang:2003ue}, which relates $U(N)$ invariant contractions under Young diagram transposition and $N\to-N$. Negative dimensionality relations have previously been applied to $U(N)$ and $SU(N)$ Wilson loops \cite{Fiol:2018yuc}, and related $SO(N)$/$Sp(N)$ relations appear in free field correlators \cite{Caputa:2013vla}.\footnote{See also \cite{Artico:2026mfc} for recent uses of negative dimensionality to theories with $O(N)$ flavor group.} Here we show that the same structure relates the giant and dual giant heavy sectors and, through Lagrangian insertion, continues to hold for separated point perturbative integrands at every loop order.

The negative rank relation also constrains possible structures beyond the correlators considered explicitly here. Maximal giant graviton correlators exhibit a partially broken higher dimensional conformal symmetry at strong coupling and also at the integrand level at weak coupling \cite{Chen:2025yxg,Chen:2026ium}. It is not known whether this structure persists for general heavy charge $M$. If it does, the negative rank relation would imply the corresponding structure for dual giant integrands. There is also an indication at finite coupling: at leading planar order, the integrated sphere and dual giant correlators of \cite{Brown:2025huy} are related by the analytic continuation $M/N\to-M/N$, precisely the continuation induced by $N\to-N$ at fixed $M$. Whether this has a finite $N$ counterpart is a sharp test of how far the negative rank relation extends beyond the perturbative integrand.

The rest of the paper is organized as follows. In Section~\ref{sec:ops}, we introduce the $\frac{1}{2}$-BPS operators and their Young diagram description. In Section~\ref{sec:omega}, we derive the negative rank relation in the free theory and discuss its connection to the negative dimensionality theorem. In Section~\ref{sec:loops}, we extend the relation to separated point loop integrands using Lagrangian insertion and discuss its implications for hidden conformal symmetry. Section~\ref{sec:checks} gives explicit checks and applies the relation to partially contracted giant gravitons. Finally, in Section~\ref{sec:finite}, we discuss the continuation of the giant and dual giant geometries at the level of the metric and the finite coupling leading planar integrated relation.

\section{$\frac{1}{2}$-BPS operators, Young diagrams and correlators}
\label{sec:ops}

\subsection{$\frac{1}{2}$-BPS operators and Young diagrams}

Let us start by working in $\mathcal N=4$ SYM with gauge group $U(N)$, leaving the extension to  $SU(N)$ to Section~\ref{sec:su}. The theory contains six real adjoint scalar fields, which we contract with a six-dimensional null R symmetry polarization vector $y$
\begin{equation}
  \Phi(x,y)\equiv y\mathbin{\cdot}\Phi(x)\,, \qquad y\mathbin{\cdot}y=0 \,.
\end{equation}
The scalar field is an adjoint matrix in color space, with components $\Phi^i{}_j$. Gauge invariant products of $p$ such fields are $\frac{1}{2}$-BPS because the null vectors automatically project the R symmetry representation to $[0,p,0]$, and they have protected conformal dimension $p$. Their color structures are conveniently labeled by permutations $\sigma\in S_p$
\begin{equation}
  \tr_\sigma\Phi \equiv \Phi^{i_1}{}_{i_{\sigma(1)}}\, \Phi^{i_2}{}_{i_{\sigma(2)}}\cdots \Phi^{i_p}{}_{i_{\sigma(p)}} \,.
\end{equation}
Every permutation has a unique decomposition into disjoint cycles. The index contractions close one color trace for each cycle, giving
\begin{equation}
  \tr_\sigma\Phi = \prod_{c\in\operatorname{cyc}(\sigma)} \tr \Phi^{|c|}\,.
\end{equation}
Here $|c|$ is the length of the cycle $c$. For later convenience, we denote the number of cycles by $C(\sigma)$, which is also the number of color traces in $\tr_\sigma\Phi$.

\vspace{0.5cm}
\noindent{\bf Young diagrams and Schur polynomials.} A more useful basis is obtained by organizing the operators into irreducible representations $R$ of $S_p$, labeled by partitions
\begin{equation}
  R=[r_1,\ldots,r_k]\vdash p\,, \qquad r_1\geq\cdots\geq r_k>0\,, \qquad \sum_{i=1}^kr_i=p \,.
\end{equation}
The partition $R$ corresponds to a Young diagram whose $i$-th row contains $r_i$ boxes. The corresponding Schur basis of charge $p$ $\frac{1}{2}$-BPS operators is defined by \cite{Corley:2001zk}
\begin{equation}
  \chi_R(\Phi) = \frac{1}{p!} \sum_{\sigma\in S_p} \chi_R(\sigma)\, \tr_\sigma\Phi\,, \label{eq:schur}
\end{equation}
where $\chi_R(\sigma)$ is the $S_p$ character in the representation $R$, and $\chi_R(\Phi)$ is the Schur polynomial associated with $R$.

\vspace{0.5cm}
\noindent{\bf Giant and dual giant gravitons.} A particular class of operators which we are interested in is that of giant and dual giant graviton operators. They take simple forms in the Schur basis, and are represented by one-column and one-row Young diagrams respectively
\begin{equation}
  \begin{aligned}
  \mathcal D_M &= \chi_{[1^M]}(\Phi) = \frac{1}{M!} \sum_{\sigma\in S_M} \operatorname{sgn}(\sigma)\, \tr_\sigma\Phi\,, \\
  \widetilde{\mathcal D}_M &= \chi_{[M]}(\Phi) = \frac{1}{M!} \sum_{\sigma\in S_M} \tr_\sigma\Phi\,.
  \end{aligned}
\end{equation}
The column representation $[1^M]$ is the sign representation of $S_M$, while the row representation $[M]$ is trivial. At finite $N$, a $U(N)$ Schur polynomial vanishes when its Young diagram has more than $N$ rows. Therefore, the giant operator has the physical charge range $0< M\leq N$, whereas the dual giant has no corresponding row length bound \cite{Corley:2001zk}.

\vspace{0.5cm}
\noindent{\bf Single-particle operators.} The other class of operators which we will consider is that of single-particle operators. They have a leading single-trace term together with higher-trace corrections
\begin{equation}
  \mathcal O_p = \tr \Phi^p +\text{(multitrace corrections)}\,,
\end{equation}
so that the single-particle operators are orthogonal to the multiparticle ones \cite{Aprile:2020uxk}. For fixed $p$ and large $N$, these corrections are suppressed by powers of $1/N$. The expansion of $\mathcal O_p$ in the Schur basis is supported on the charge-$p$ hook diagrams $R_k^p=[p-k+1,1^{k-1}]$, with $1\leq k\leq p$ \cite{Aprile:2020uxk}. In the convention above, where the coefficient of the leading single trace is one, the exact expansion is
\begin{equation}
  \begin{aligned}
  \mathcal O_p &=\sum_{k=1}^pc_k(p,N)\,\chi_{R_k^p}(\Phi)\,, \\
  c_k(p,N) &=\frac{p(p-1)(-1)^{k-1}(N-p+1)_{p-k} (N+p-k+1)_{k-1}}{(N)_p-(N+1-p)_p}\,,
  \end{aligned}
  \label{eq:hook}
\end{equation}
where $(z)_r=\prod_{m=0}^{r-1}(z+m)$ is the rising Pochhammer symbol, with $(z)_0=1$. Let us also mention that for integer $N$ the operator $\mathcal O_p$ vanishes for $p>N$. This is the field theory realization of the stringy exclusion principle in the single-particle sector \cite{Aprile:2020uxk}.\footnote{This does not mean that $\mathcal O_N$ is identical to $\mathcal D_N$ at finite $N$. It becomes proportional to $\mathcal D_N$ only in the large $N$ limit with $p=N\to \infty$.}

\subsection{Correlators and Wick contractions}

The results that we will derive are applicable to $n$-point $\frac{1}{2}$-BPS correlators of arbitrary Schur operators. But here for concreteness we will illustrate them with four-point functions where the heavy operators are giant or dual giant operators and the light insertions are protected single-particle operators. Explicitly, the correlators are
\begin{equation}
  \begin{aligned}
  G(N;M,p,q) &\equiv \left\langle \mathcal D_M(1)\mathcal D_M(2) \mathcal O_p(3)\mathcal O_q(4) \right\rangle\,, \\
  \widetilde G(N;M,p,q) &\equiv \left\langle \widetilde{\mathcal D}_M(1) \widetilde{\mathcal D}_M(2) \mathcal O_p(3)\mathcal O_q(4) \right\rangle\,.
  \end{aligned}
\end{equation}
For the moment, we restrict the discussion to the free theory level. We will introduce the notation where we add a superscript $(0)$ to the correlators and use $\langle\cdots\rangle_0$ in the expectation value to denote evaluation by free field Wick contractions. The elementary scalar contraction is
\begin{equation}
  \contraction{(}{\Phi}{_i)^a{}_b\,(}{\Phi} (\Phi_i)^a{}_b\,(\Phi_j)^c{}_d = \frac{d_{ij}}{4\pi^2}\, \delta^a{}_d\,\delta^c{}_b\,,
\end{equation}
where
\begin{equation}
  d_{ij}\equiv \frac{y_i\mathbin{\cdot}y_j}{x_{ij}^2}\,, \qquad x_{ij}^2\equiv(x_i-x_j)^2\,.
\end{equation}

\section{The trace reflection and the negative rank relation}
\label{sec:omega}

\subsection{Trace reflection}

We start this section by introducing an operator which allows us to relate different types of $\frac{1}{2}$-BPS operators. We will first motivate it by looking at giant and dual giant gravitons, before talking about general Schur polynomials and then focusing on the single-particle operators.

\vspace{0.5cm}
\noindent{\bf Giant and dual giant gravitons.} The giant and dual giant operators introduced in the previous section are closely related already in their definitions. The one-column diagram of $\mathcal D_M$ and the one-row diagram of $\widetilde{\mathcal D}_M$ are transposes of one another, suggesting a close relation between their correlators. To make this relation precise, let us start by defining the following trace reflection operator $\omega$
\begin{equation}
  \begin{aligned}
  \omega\!\left[\tr W\right]&=-\tr W\,,\qquad \omega\!\left[   \tr W_1\cdots\tr W_n \right] &=(-1)^n\tr W_1\cdots\tr W_n\,.
  \end{aligned}
\end{equation}
For consistency with the trace of the identity matrix $\tr \mathbf 1=\delta^i{}_i=N$, we also define
\begin{equation}
  \omega[\tr \mathbf 1] =\omega[N]=-N\,.
\end{equation}
Thus $\omega$ changes the sign of each explicit color trace and sends $N\to-N$, while leaving the matrix words inside the traces unchanged.\footnote{The operation is defined on trace polynomials before finite $N$ trace identities are imposed. For example, at $N=1$, $\tr \Phi^2=(\tr \Phi)^2$ since $\Phi$ is just a number, so equivalent representatives can contain different numbers of traces and $\omega$ would otherwise be ill defined.}

The relation between the heavy operators is exposed directly by their determinant generating functions. As formal power series in $t$, we have
\begin{equation}
  \begin{aligned}
  \det(1+t\Phi) &=   \exp\!\left[\tr \log(1+t\Phi)\right] \\
  &= \exp\!\left[     \sum_{r\geq1}\frac{(-1)^{r-1}t^r}{r}      \tr \Phi^r    \right]=\sum_{M\geq0}t^M\mathcal D_M\,, \\
  \det(1-t\Phi)^{-1} &=\exp\!\left[-\tr \log(1-t\Phi)\right] \\
  &=\exp\!\left[\sum_{r\geq1}\frac{t^r}{r}\tr \Phi^r\right]=\sum_{M\geq0}t^M\widetilde{\mathcal D}_M\,,
  \end{aligned}
\end{equation}
and the giant and dual giant graviton operators correspond to the coefficients of these series. Applying $\omega$ to the first exponential reverses the sign of every trace, and produces the dual giant generating function with $t$ replaced by $-t$
\begin{equation}
  \omega\!\left[\det(1+t\Phi)\right] =\det(1+t\Phi)^{-1} =\sum_{M\geq0}(-t)^M\widetilde{\mathcal D}_M\,.
\end{equation}
Comparing coefficients then gives
\begin{equation}
  \omega\!\left[\mathcal D_M\right]=(-1)^M\widetilde{\mathcal D}_M\,.
\end{equation}

\vspace{0.5cm}
\noindent{\bf Young diagram transposition.} To establish the action of $\omega$ on generic Young diagrams, we use the permutation formula~\eqref{eq:schur} for the Schur polynomial. Let $C(\sigma)$ denote the number of cycles of $\sigma\in S_p$. Since each cycle produces one explicit color trace, we have
\begin{equation}
  \omega\!\left[\tr_\sigma\Phi\right]=(-1)^{C(\sigma)} \tr_\sigma\Phi\,,
\end{equation}
where the factor $(-1)^{C(\sigma)}$ is related to the signature of $\sigma$ by
\begin{equation}
  \operatorname{sgn}(\sigma)= (-1)^{p - C(\sigma)}\,.
\end{equation}
We also have the following relation between the character associated with a Young diagram $R$ and that of its transpose $R^T$
\begin{equation}
  \chi_{R^T}(\sigma) =\operatorname{sgn}(\sigma)\chi_R(\sigma)\,. \label{eq:chartrans}
\end{equation}
To see why this relation holds, we recall that the representation associated with a Young diagram is constructed by symmetrizing indices along each row and antisymmetrizing them along each column. Tensoring with the sign representation multiplies every permutation by its parity. It therefore turns row symmetrization into antisymmetrization and column antisymmetrization into symmetrization, exchanging the roles of rows and columns. Thus $\operatorname{sgn}\otimes R$ is labeled by $R^T$. Since the sign representation is one dimensional, its character multiplies $\chi_R(\sigma)$ by $\operatorname{sgn}(\sigma)$ and gives equation~\eqref{eq:chartrans}. Inserting these relations into the permutation formula for the Schur polynomial gives
\begin{equation}
  \begin{aligned}
  \omega\!\left[\chi_R(\Phi)\right] &=\frac{1}{p!}\sum_{\sigma\in S_p}   (-1)^{C(\sigma)}    \chi_R(\sigma)\,  \tr_\sigma\Phi \\
  &= (-1)^p\frac{1}{p!} \sum_{\sigma\in S_p}  \chi_{R^T}(\sigma)\, \tr_\sigma\Phi=(-1)^p\chi_{R^T}(\Phi)\,.
  \end{aligned}
  \label{eq:schurtrans}
\end{equation}
Equation~\eqref{eq:schurtrans} is the general form of the exchange seen in the determinant example. Since the one-column and one-row diagrams are transposes of one another, we directly obtain
\begin{equation}
  \omega\!\left[\mathcal D_M\right]=(-1)^M\widetilde{\mathcal D}_M\,, \qquad \omega\!\left[\widetilde{\mathcal D}_M\right]=(-1)^M\mathcal D_M\,.
\end{equation}

\vspace{0.5cm}
\noindent{\bf Single-particle operators.} Let us now consider the action on the single-particle operator $\mathcal O_p$ whose exact hook expansion was given in equation~\eqref{eq:hook}. Equation~\eqref{eq:schurtrans} shows that the trace sign part of $\omega$ sends each hook to its transposed hook, with the common sign $(-1)^p$. Hook transposition acts as $(R_k^p)^T=R_{p-k+1}^p$. The rank reflection part of $\omega$ acts on the coefficients, giving
\begin{equation}
  c_{p-k+1}(p,N)=(-1)^{p-1}c_k(p,-N)\,.
\end{equation}
Combining the two parts of $\omega$, we find
\begin{equation}
  \omega\!\left[\mathcal O_p\right] = (-1)^p\sum_{k=1}^pc_k(p,-N)\,\chi_{R_{p-k+1}^p}(\Phi)=-\mathcal O_p\,.
\end{equation}
Thus $\omega$ acts on the single-particle operator with eigenvalue $-1$.

\subsection{Exact $N$ Wick contractions}
\label{sec:wick}

The operator identities above do not yet imply a relation between correlators. Already at the free level, Wick contractions of free fields close color index loops and generate additional $N$ dependence. We first show that this dependence transforms correctly under $\omega$. The extension to loop integrands will be given in Section~\ref{sec:loops}.

The key argument below depends only on how the color indices are contracted. Therefore, we will suppress the spacetime and $R$ symmetry dependence to avoid overloading the expressions with unnecessary information. We say a free correlator has total charge $2E$ if it is a sum of Wick contractions of trace monomials containing $2E$ scalar fields $\Phi$. We label these scalar fields by $1,\ldots,2E$. The trace structure is encoded by a permutation $\alpha\in S_{2E}$, whose $C(\alpha)$ cycles are the explicit traces
\begin{equation}
  F_\alpha=\Phi^{i_1}{}_{i_{\alpha(1)}}\cdots\Phi^{i_{2E}}{}_{i_{\alpha(2E)}}=\tr W_1\cdots\tr W_{C(\alpha)}\,.
\end{equation}
A Wick contraction selects two scalar fields, say the $m$-th and $n$-th, and replaces them by the propagator. Suppressing its spacetime and $R$ symmetry factor, the Wick contraction has the color structure
\begin{equation}
  \contraction{}{\Phi}{^{i_m}{}_{i_{\alpha(m)}}}{\Phi} \Phi^{i_m}{}_{i_{\alpha(m)}}\Phi^{i_n}{}_{i_{\alpha(n)}}\propto\delta^{i_m}{}_{i_{\alpha(n)}}\delta^{i_n}{}_{i_{\alpha(m)}}\,.
\end{equation}
We can record a specific Wick pairing by the involution $\beta=(m_1\,n_1)\cdots(m_E\,n_E)$. For this pairing, the color factor is determined by the standard permutation cycle count \cite{Corley:2001zk}
\begin{equation}
  F_{\alpha,\beta} = \delta^{i_1}{}_{i_{\alpha(\beta(1))}} \cdots \delta^{i_{2E}}{}_{i_{\alpha(\beta(2E))}}=N^{C(\alpha\beta)}\,.
\end{equation}
We now apply $\omega$ to the trace structure $\alpha$ with the Wick pairing $\beta$. Its action on the explicit traces contributes $(-1)^{C(\alpha)}$, while evaluation of the closed color loops at $-N$ contributes the factor $(-1)^{C(\alpha\beta)}$ instead. Therefore, each term with this Wick pairing acquires the sign
\begin{equation}
  (-1)^{C(\alpha)}(-1)^{C(\alpha\beta)} = \operatorname{sgn}(\alpha)\operatorname{sgn}(\alpha\beta) = \operatorname{sgn}(\beta) = (-1)^E\,.
\end{equation}
This sign depends only on the total charge $2E$ of the correlator. Consequently, for a free correlator of trace polynomial operators $F_1,\ldots,F_n$ with total charge $2E$, summing over all Wick pairings gives
\begin{equation}
  \left\langle F_1\cdots F_n\right\rangle_{U(N)}^{(0)} =(-1)^E\left\langle\omega[F_1]\cdots\omega[F_n]\right\rangle_{U(-N)}^{(0)}\,. \label{eq:wick}
\end{equation}
Here $\langle\cdots\rangle_{U(-N)}^{(0)}$ denotes formal evaluation of the reflected expression at rank $-N$, including its explicit rank-dependent coefficients, so every closed color index loop contributes $\delta^i{}_i=-N$. The rank reflection inside $\omega$ compensates this evaluation on such coefficients in the fixed-pairing identity. We will refer to equation~\eqref{eq:wick} as the negative rank relation. It is a trace basis realization of the classical negative dimensionality theorem \cite{Cvitanovic:1982bq,Elvang:2003ue}, as will be made explicit in Section~\ref{sec:negdim}. As a direct application, for the HHLL operators the total number of fields is $2M+p+q=2E$, hence $E=M+(p+q)/2$. Applying $\omega$ to the external operators gives
\begin{equation}
  \widetilde G^{(0)}(N;M,p,q) = (-1)^{M+(p+q)/2} G^{(0)}(-N;M,p,q)\,. \label{eq:hhll}
\end{equation}

It is also common to rescale the external operators, in particular to give $\mathcal O_p$ a unit two-point function. Such a rescaling preserves the relation because the normalization coefficients are themselves free two-point functions. Writing
\begin{equation}
  \left\langle \mathcal O_p(1)\mathcal O_p(2) \right\rangle = \mathcal N_{p,N} \left(\frac{d_{12}}{4\pi^2}\right)^p\,,
\end{equation}
and denoting the analogous giant and dual giant coefficients by $\mathcal N_{M,N}$ and $\widetilde{\mathcal N}_{M,N}$, the same pairing argument gives
\begin{equation}
  \widetilde{\mathcal N}_{M,N} = (-1)^M\mathcal N_{M,-N}\,, \qquad \mathcal N_{p,N} = (-1)^p\mathcal N_{p,-N}\,.
\end{equation}
With consistent signs for the square roots, these phases cancel the $(-1)^E$ in equation~\eqref{eq:hhll} when all external operators are unit normalized.

The pairing argument does not rely on the HHLL setup and applies to any number of Schur polynomial insertions. It also extends to adjoint fermions, gauge fields and ghosts, whose propagators have the same color structure.

\subsection{Traceless projection and the \texorpdfstring{$SU(N)$}{SU(N)} relation}
\label{sec:su}

The $SU(N)$ relation follows directly from the $U(N)$ pairing theorem once the traceless adjoint field is realized inside the $U(N)$ theory. Let us define the traceless part of the $U(N)$ matrix by \cite{Aprile:2020uxk}
\begin{equation}
  \widehat\Phi=\Phi-\left(\frac{\tr \Phi}{N}\right)\mathbf 1\,.
\end{equation}
Subtracting this singlet component from both fields removes precisely the trace part of the canonical $U(N)$ contraction, giving
\begin{equation}
  \left. \contraction{(}{\widehat\Phi}{_i)^a{}_b\,(}{\widehat\Phi} (\widehat\Phi_i)^a{}_b\,(\widehat\Phi_j)^c{}_d \right|_{U(N)} = \frac{d_{ij}}{4\pi^2} \left( \delta^a{}_d\delta^c{}_b -\frac{1}{N}\delta^a{}_b\delta^c{}_d \right) = \left. \contraction{(}{\Phi}{_i)^a{}_b\,(}{\Phi} (\Phi_i)^a{}_b\,(\Phi_j)^c{}_d \right|_{SU(N)}\,. \label{eq:suprop}
\end{equation}
Here $\Phi$ on the right is the elementary traceless adjoint field of the $SU(N)$ theory. The projected $U(N)$ field and the elementary $SU(N)$ field therefore have identical propagators. So for every polynomial $F$ we have
\begin{equation}
  \langle F[\widehat\Phi]\rangle_{U(N)}^{(0)} = \left\langle F[\Phi]\right\rangle_{SU(N)}^{(0)}\,. \label{eq:suproj}
\end{equation}
Note that the projector is also compatible with the trace reflection. The unit matrix ${\bf 1}$ is left unchanged and the signs from the trace and rank parts of $\omega$ cancel. Therefore, we get
\begin{equation}
  \omega[\widehat\Phi]=\Phi-\frac{\tr \Phi}{N}\mathbf 1=\widehat\Phi\,.
\end{equation}
We can then apply~\eqref{eq:wick} to the projected $U(N)$ fields and use~\eqref{eq:suproj} on both sides. It follows
\begin{equation}
  \left\langle F_1\cdots F_n\right\rangle_{SU(N)}^{(0)} = (-1)^E \left\langle \omega[F_1]\cdots\omega[F_n] \right\rangle_{SU(-N)}^{(0)}\,, \label{eq:suwick}
\end{equation}
and we similarly also get the $SU(N)$ counterpart of equation~\eqref{eq:hhll} with the same factor.

\subsection{Relation to the negative dimensionality theorem}
\label{sec:negdim}
The relation \eqref{eq:wick} is closely connected to the classical negative dimensionality theorem \cite{Cvitanovic:1982bq,Elvang:2003ue}. The theorem states that, for a $U(N)$ scalar whose color indices are fully contracted using $U(N)$ Young projectors, exchanging symmetrizers and antisymmetrizers is equivalent to evaluating the contractions in $U(-N)$.\footnote{Strictly speaking, the result is up to an overall factor $(-1)^{|R|}$ where $|R|$ is the size of the corresponding Young diagram. In our case, the size is equal to the number of fields, which is always even.}

To see the connection, we rewrite each Wick contraction in terms of symmetric and antisymmetric projectors
\begin{equation}
  \begin{aligned}
  \contraction{(}{\Phi}{_i)^a{}_b\,(}{\Phi} (\Phi_i)^a{}_b\,(\Phi_j)^c{}_d \propto \delta^a{}_d\delta^c{}_b = \frac{1}{2}\left( \delta^a{}_b\delta^c{}_d +\delta^a{}_d\delta^c{}_b \right) - \frac{1}{2}\left( \delta^a{}_b\delta^c{}_d -\delta^a{}_d\delta^c{}_b \right)\,.
  \end{aligned}
\end{equation}
The resulting action on the color indices is therefore a symmetric projector minus an antisymmetric projector. The full free correlator can thus be viewed as a $U(N)$ scalar in the sense of \cite{Elvang:2003ue}. Under the negative dimensionality theorem, exchanging symmetrizers and antisymmetrizers maps the $U(N)$ contractions to their $U(-N)$ counterparts. Each of the $E$ Wick pairs acquires a minus sign under this exchange, producing the factor $(-1)^E$ in equation~\eqref{eq:wick}.

\section{Lagrangian insertion and loop integrands}
\label{sec:loops}

In this section, we extend the free theory argument to include perturbative interactions by using Lagrangian insertion. In this formalism, an $L$-loop correlator is obtained from a Born-level correlator with $L$ insertions of the chiral on-shell Lagrangian \cite{Eden:2011yp}. The resulting integrand can again be computed by Wick contractions, so the argument of the previous section still applies. The additional Lagrangian insertions, however, require some extra care, as we discuss below.

\subsection{Lagrangian insertion}

We adopt the conventions of Appendix~A of \cite{He:2024cej} and work in Lorentzian signature in this section.\footnote{Our previous argument depends only on color contractions and is independent of the signature.} The action of $\mathcal N=4$ SYM is
\begin{equation}
  S=\int d^4x\,L_{\rm full}\,,
\end{equation}
with
\begin{equation}
  \begin{aligned}
  L_{\rm full}=\tr \Bigl( &-\frac{1}{2}F_{\alpha\beta}F^{\alpha\beta} -\frac{1}{4}\Phi^{AB}[D_{\alpha\dot\alpha}, [D^{\dot\alpha\alpha},\Phi_{AB}]] +\frac{i}{2}\bar\Psi_{A\dot\alpha} [D^{\dot\alpha\alpha},\Psi^A_\alpha] \\
  &-\frac{3i}{2}[D^{\dot\alpha\alpha},\bar\Psi_{A\dot\alpha}] \Psi^A_\alpha +\frac{g_{YM}^2}{8} [\Phi^{AB},\Phi^{CD}][\Phi_{AB},\Phi_{CD}] \\
  &-\sqrt{2}g_{YM} \Psi^{A\alpha}[\Phi_{AB},\Psi^B_\alpha] +\sqrt{2}g_{YM} \bar\Psi_{A\dot\alpha}[\Phi^{AB},\bar\Psi_B^{\dot\alpha}]\Bigr)\,,
  \end{aligned}
\end{equation}
and
\begin{equation}
  D_\mu=\partial_\mu-i g_{YM}A_\mu\,, \qquad F_{\mu\nu}=i g_{YM}^{-1}[D_\mu,D_\nu]\,.
\end{equation}
In this convention, the kinetic terms are independent of the coupling, while interactions carry powers of $g_{YM}$. The full correlators therefore have the perturbative expansions
\begin{equation}
  \begin{aligned}
  G(N,g_{YM}^2;M,p,q) &= \sum_{L=0}^\infty \bigl(g_{YM}^2\bigr)^L G^{(L)}(N;M,p,q)\,, \\
  \widetilde G(N,g_{YM}^2;M,p,q) &= \sum_{L=0}^\infty \bigl(g_{YM}^2\bigr)^L \widetilde G^{(L)}(N;M,p,q)\,.
  \end{aligned}
\end{equation}
Here $G^{(L)}$ and $\widetilde G^{(L)}$ are the $L$-loop correlators in the perturbative expansion. Applying $(\partial/\partial g_{YM}^2)^L/L!$ to the full correlator and then taking its $g_{YM}^0$ part extracts the $L$-loop correlator. Ultimately, the Lagrangian insertion method states that each derivative effectively inserts one copy of the chiral on-shell Lagrangian $\mathcal L_{\rm chiral}(x_r)/g_{YM}^2$ \cite{Eden:2011yp,He:2024cej},\footnote{At mutually separated points the coupling deformation may be represented by the chiral on-shell Lagrangian. Differences involving equations of motion contribute only contact terms.} where
\begin{equation}
  \mathcal L_{\rm chiral} = \tr \!\left( -\frac{1}{2}F_{\alpha\beta}F^{\alpha\beta} +\sqrt{2}g_{YM} \Psi^{A\alpha}[\Phi_{AB},\Psi^B_\alpha] -\frac{g_{YM}^2}{8} [\Phi^{AB},\Phi^{CD}][\Phi_{AB},\Phi_{CD}] \right)\,.
\end{equation}
The correlator with these insertions is evaluated at Born level, giving
\begin{equation}
  G^{(L)} = \frac{(-i)^L}{L!} \int\prod_{r=5}^{4+L}d^4x_r\, \mathcal I^{(L)}\,, \qquad \widetilde G^{(L)} = \frac{(-i)^L}{L!} \int\prod_{r=5}^{4+L}d^4x_r\, \widetilde{\mathcal I}^{(L)}\,.
\end{equation}
The integrands are
\begin{equation}
  \begin{aligned}
  \mathcal I^{(L)}(N;M,p,q) &= \left\langle \mathcal D_M(1)\mathcal D_M(2) \mathcal O_p(3)\mathcal O_q(4) \prod_{r=5}^{4+L} \frac{\mathcal L_{\rm chiral}(x_r)}{g_{YM}^2} \right\rangle^{(0)}\,, \\
  \widetilde{\mathcal I}^{(L)}(N;M,p,q) &= \left\langle \widetilde{\mathcal D}_M(1) \widetilde{\mathcal D}_M(2) \mathcal O_p(3)\mathcal O_q(4) \prod_{r=5}^{4+L} \frac{\mathcal L_{\rm chiral}(x_r)}{g_{YM}^2} \right\rangle^{(0)}\,,
  \end{aligned}
\end{equation}
where the superscript $(0)$ denotes the $g_{YM}^0$ part of the full interacting correlator. For the correlators considered in the previous sections, this part coincides with the direct free field Wick contraction $\langle\cdots\rangle_0$. On the other hand, a correlator with $\mathcal L_{\rm chiral}/g_{YM}^2$ insertions contains explicit inverse powers of the coupling, so its $g_{YM}^0$ part also receives contributions from the ordinary interaction vertices in the action \cite{Eden:2011yp}. To make the distinction explicit, we decompose $S$ and $\mathcal L_{\rm chiral}$ according to powers of the coupling:
\begin{equation}
  S=S_0+g_{YM}S_1+g_{YM}^2S_2\,, \qquad \mathcal L_{\rm chiral}=\mathcal L_0+g_{YM}\mathcal L_1 +g_{YM}^2\mathcal L_2\,.
\end{equation}
For $\mathcal F=\mathcal D_M(1)\mathcal D_M(2)\mathcal O_p(3)\mathcal O_q(4)$ with one inserted chiral Lagrangian, the $g_{YM}^0$ part is:\footnote{There is no term with an $S_0$ insertion because the quadratic action $S_0$ defines the kinetic term rather than an interaction vertex.}
\begin{equation}
  \begin{aligned}
  \left\langle \mathcal F\frac{\mathcal L_{\rm chiral}(x)}{g_{YM}^2} \right\rangle^{(0)} ={}& \left\langle \mathcal F\mathcal L_2(x)\right\rangle_0 +i\left\langle \mathcal F\mathcal L_1(x)S_1\right\rangle_0 +i\left\langle \mathcal F\mathcal L_0(x)S_2\right\rangle_0 \\
  &-\frac{1}{2} \left\langle \mathcal F\mathcal L_0(x)S_1^2\right\rangle_0\,.
  \end{aligned}
\end{equation}
Here $\langle\cdots\rangle_0$ means the free field Wick contraction. The ordinary vertex integrations and the subtractions required by the normalized path integral are understood. This example shows why the $g_{YM}^0$ part $\langle\cdots\rangle^{(0)}$ of an inserted correlator is not, in general, the direct free field Wick contraction $\langle\cdots\rangle_0$ of the displayed operators.

\subsection{Negative rank relation for loop integrands}
\label{sec:looprel}

We now extend the argument of Section~\ref{sec:wick} to establish the negative rank relation for $L$-loop integrands. We apply it to each free field Wick pairing, now including the additional chiral Lagrangian insertions and interaction vertices. Let $F_1,\ldots,F_n$ be trace polynomial external insertions of total scalar charge $2E$, using the notation of Section~\ref{sec:wick}. For the HHLL product, we have
\begin{equation}
  2E=2M+p+q\,.
\end{equation}
Let $P$ be the total number of Wick pairs in one term with $L$ Lagrangian insertions and $V$ interaction vertices. To determine $P$, we first note that expanding the field strength organizes the terms in the chiral Lagrangian according to their powers of the coupling
\begin{equation}
  \mathcal L_{\rm chiral} \sim \tr (\partial A\,\partial A) +g_{YM}\tr \bigl([A,A]\,\partial A+\Psi[\Phi,\Psi]\bigr) +g_{YM}^2\tr \bigl([A,A]^2+[\Phi,\Phi]^2\bigr)\,.
\end{equation}
Thus a term proportional to $g_{YM}^k$ contains $k+2$ elementary fields. The same counting holds for the terms in the action \cite{He:2024cej}.\footnote{In the computation one needs to fix the gauge and include ghosts in the action. In a linear covariant gauge, the ghost Lagrangian is $2\tr [(\partial^\mu\bar c)(\partial_\mu c-i g_{YM}[A_\mu,c])]$ \cite{Capri:2014tta}, which obeys the same count.} Suppose the $L$ chiral Lagrangian insertions and the $V$ interaction vertices contribute powers $g_{YM}^{k_a}$, with $a=1,\ldots,L+V$. Each Lagrangian insertion contributes an extra factor $g_{YM}^{-2}$, so selecting the $g_{YM}^0$ part requires
\begin{equation}
  \sum_{a=1}^{L+V}k_a=2L\,.
\end{equation}
The total number of fields to be Wick contracted is therefore
\begin{equation}
  2P = 2E+\sum_{a=1}^{L+V}(k_a+2) = 2E+4L+2V\,, \qquad P=E+2L+V\,.
\end{equation}
For each Wick pairing, the argument of Section~\ref{sec:wick} applied to all traces gives $(-1)^P$. We use $\left\langle F_1\cdots F_n\right\rangle^{(L)}$ to denote the {\it integrand} at $L$-loop of the correlator of $F_i$'s. The desired comparison, however, is between correlators in which $\omega$ acts only on the external operator product
\begin{equation}
  \left\langle \omega[F_1]\cdots\omega[F_n] \right\rangle_{U(-N)}^{(L)} = \left\langle \omega[F_1]\cdots\omega[F_n] \prod_{r=n+1}^{n+L} \frac{\mathcal L_{\rm chiral}(x_r)}{g_{YM}^2} \right\rangle_{U(-N)}^{(0)}\,,
\end{equation}
rather than a correlator in which the explicit traces in $\mathcal L_{\rm chiral}$ and $S$ are also reflected. Since each of the $L$ insertions and $V$ vertices is a single trace, leaving these traces unreflected contributes an additional factor $(-1)^{L+V}$. The total sign is
\begin{equation}
  (-1)^P(-1)^{L+V} =(-1)^{E+L}\,.
\end{equation}
Thus, at the integrand level,
\begin{equation}
  \left\langle F_1\cdots F_n\right\rangle_{U(N)}^{(L)} = (-1)^{E+L} \left\langle \omega[F_1]\cdots\omega[F_n] \right\rangle_{U(-N)}^{(L)}\,.
\end{equation}
For Schur polynomial external operators $F_i$, $\omega$ transposes every Young diagram, while the degree signs multiply to $(-1)^{2E}=1$. Hence the relation holds for any number of Schur insertions at every loop order, provided all external and insertion points are separated. For the HHLL product, it gives
\begin{equation}
  \widetilde{\mathcal I}^{(L)}(N;M,p,q) = (-1)^{E+L}\mathcal I^{(L)}(-N;M,p,q)\,.
\end{equation}
The same general argument applies to $SU(N)$, or equivalently follows from the projection established in Section~\ref{sec:su}. For the HHLL product, it gives
\begin{equation}
  \widetilde{\mathcal I}_{SU(N)}^{(L)} = (-1)^{E+L}\mathcal I_{SU(-N)}^{(L)}\,.
\end{equation}

The factor $(-1)^L$ can be absorbed by taking $g_{YM}^2 \to -g_{YM}^2$. Together with $N\to-N$, this leaves the 't Hooft coupling $\lambda=g_{YM}^2N$ fixed, so the integrand relation carries only the charge-dependent sign $(-1)^E$. Our argument applies to separated point integrands. Extending it to integrated correlators requires a compatible treatment of contact terms, regularization and operator renormalization. We will not consider these issues here and leave this extension for future work.

\subsection{Implications for hidden conformal symmetry}

Let us now make a brief comment about potential implications of our negative rank relation. A remarkable property of spherical giant graviton correlators is a partially broken higher dimensional conformal symmetry \cite{Chen:2025yxg,Chen:2026ium}. This higher dimensional structure was originally identified for four-point functions of light $\frac12$-BPS single-particle operators \cite{Caron-Huot:2018kta,Caron-Huot:2021usw,Caron-Huot:2023wdh}, and was later found to appear in a partially broken form at strong coupling for four-point functions of two maximal giant gravitons with $M=N$ and two light operators. In the defect interpretation, the correlators of different light operator weights can be packaged into a single generating function by uplifting the spacetime and R symmetry kinematics to a higher dimensional embedding space. Remarkably, the same structure persists at weak coupling at the level of Lagrangian insertion integrands: at each loop order, the integrands with different light operator weights are generated from the lowest-weight one by the same higher dimensional uplift.

It would be interesting to test whether this hidden conformal symmetry extends to giants of generic size, with $M$ treated independently of $N$. If the hidden conformal symmetry of the giant graviton integrands can be established for general $M$, then our negative rank relation immediately implies that the corresponding dual giant graviton integrands exhibit exactly the same structure. It is important that this extension be established before reflecting the rank, since the hidden structure could in principle rely on simplifications that occur only after setting $M=N$. Furthermore, if such a generic $M$ extension is established at weak coupling, we can reverse the logic of \cite{Chen:2025yxg,Chen:2026ium} and look for the partially broken hidden conformal symmetry of dual giant graviton correlators at strong coupling. We note that the latter correlators in the supergravity limit remain to be computed.

\section{Examples and applications}
\label{sec:checks}

In this section we provide several examples to verify the negative rank relation explicitly. Then we apply the relation to derive partially contracted giant graviton (PCGG) formulas for dual giants.

\subsection{Light correlators}

As a first check, we use the connected $SU(N)$ four-point correlator given in Appendix D of \cite{Aprile:2020uxk}. The relevant single-particle operators are
\begin{equation}
  \mathcal O_3=\tr \Phi^3\,,\qquad \mathcal O_4 =\tr \Phi^4-\frac{2N^2-3}{N(N^2+1)}\bigl(\tr \Phi^2\bigr)^2\,.
\end{equation}
Stripping off a numerical factor $(4\pi^2)^{-7}$ from the propagators, the connected correlator is
\begin{equation}
  \begin{aligned}
  G_{3344}(N) =&\ (4\pi^2)^7 \left\langle \mathcal O_3(1)\mathcal O_3(2)\mathcal O_4(3)\mathcal O_4(4)\right\rangle_{SU(N),\ {\rm connected}} \\
  =&\ \alpha_1\left(  d_{12}^2d_{13}d_{24}d_{34}^3+d_{12}^2d_{23}d_{14}d_{34}^3\right)+\alpha_3\left(d_{13}^2d_{12}d_{24}^2d_{34}^2+d_{23}^2d_{12}d_{14}^2d_{34}^2 \right) \\
  &+ \alpha_4\left( d_{13}^2d_{23}d_{14}d_{24}^2d_{34} +d_{23}^2d_{13}d_{14}^2d_{24}d_{34}  \right)+\alpha_7d_{12}d_{13}d_{23}d_{14}d_{24}d_{34}^2\,,
  \end{aligned}
\end{equation}
where the coefficients are given by
\begin{equation}
  \begin{aligned}
  \alpha_1 =36\frac{N^2-4}{N}R_4\,,\quad 2\alpha_3&=\alpha_4=72\frac{N(N^2-9)}{N^2+1}R_4\,,\\
  \alpha_7=72\frac{N^4-25N^2-6}{N(N^2+1)}R_4\,,\quad R_4&=4\frac{(N^2-1)(N^2-4)(N^2-9)}{N^2+1}\,.
  \end{aligned}
\end{equation}
Here $R_4$ is even under $N\to-N$, while every $\alpha_i$ is odd. Consequently, this connected $SU(N)$ correlator obeys
\begin{equation}
  G_{3344}(-N)=-G_{3344}(N)\,,
\end{equation}
and is consistent with equation~\eqref{eq:suwick}. For the two-point normalized correlator, the two $\mathcal O_3$ and two $\mathcal O_4$ insertions contribute the total external normalization denominator $R_3R_4$, with the $\mathcal O_3$ two-point coefficient
\begin{equation}
  R_3 =3\frac{(N^2-1)(N^2-4)}{N}\,.
\end{equation}
Both the unnormalized connected correlator and $R_3R_4$ are odd under $N\to-N$, so their ratio is even under rank reflection.

\subsection{HHL and HHLL correlators}
\label{sec:hhl}

Heavy-heavy-light correlators also give simple explicit examples of the negative rank relation. Since the kinematic dependence is fixed by symmetry, we suppress it below. We first consider the extremal correlators of two giants and one single trace operator computed at finite $N$ in \cite{Bissi:2011dc}. In our notation, the result reads
\begin{equation}
  \left\langle \widetilde{\mathcal D}_M\, \widetilde{\mathcal D}_{M-p}\, \tr \Phi^p \right\rangle_{U(N)} = \prod_{j=1}^M(N-1+j)\,,
\end{equation}
while
\begin{equation}
  \left\langle \mathcal D_M\, \mathcal D_{M-p}\, \tr \Phi^p \right\rangle_{U(N)} = (-1)^{p-1} \prod_{j=1}^M(N-j+1)\,.
\end{equation}
Hence
\begin{equation}
  \left\langle \widetilde{\mathcal D}_M\, \widetilde{\mathcal D}_{M-p}\, \tr \Phi^p \right\rangle_{U(N)} = (-1)^{M+p-1} \left\langle \mathcal D_M\, \mathcal D_{M-p}\, \tr \Phi^p \right\rangle_{U(-N)}\,.
\end{equation}
A more nontrivial example is provided by the nonextremal HHL correlators at leading large $N$ in \cite{Yang:2021kot}. Their result for two equal sphere giants is
\begin{equation}
  \left\langle \mathcal D_M\, \mathcal D_M\, \tr \Phi^p \right\rangle_{U(N)} = -\frac{i^p+(-i)^p}{2\sqrt p} \left[ P_{p/2}\!\left(2\frac{M}{N}-1\right) + P_{p/2-1}\!\left(2\frac{M}{N}-1\right) \right]\,,
\end{equation}
and the corresponding dual giant result is
\begin{equation}
  \left\langle \widetilde{\mathcal D}_M\, \widetilde{\mathcal D}_M\, \tr \Phi^p \right\rangle_{U(N)} = \frac{1+(-1)^p}{2\sqrt p} \left[ P_{p/2}\!\left(1+2\frac{M}{N}\right) - P_{p/2-1}\!\left(1+2\frac{M}{N}\right) \right]\,,
\end{equation}
where $P_n(x)$ is the Legendre polynomial. These correlators are nonzero only for even $p$. Using
\begin{equation}
  P_n(-x)=(-1)^nP_n(x)\,,
\end{equation}
the two expressions are mapped into one another by $N \to -N$ up to a minus sign.

We next turn to an interacting HHLL example through three loops in \cite{He:2026ios}. They study the leading large $N$ integrand of two giants or dual giants, with two light operators $\mathcal O_2$, in an expansion in the 't Hooft coupling. In their results, they introduce the parameter $\alpha$
\begin{equation}
  \alpha=\frac{M}{N}\in(0,1] \quad\text{for a giant}\,, \qquad \alpha=-\frac{M}{N}<0 \quad\text{for a dual giant}\,,
\end{equation}
and show that the integrands for both giants and dual giants are described by the same function of the signed parameter $\alpha$. Their results for the integrands $F_{4+L}$, up to three loops, take the form
\begin{align}
  F_5={}&f^{(1)}\,,\\
  F_6={}&f_2^{(2)}-\alpha f_1^{(2)}\,,\\
  F_7={}& -\alpha f_{1,1}^{(3)} +f_{1,2}^{(3)} +f_{1,3}^{(3)} +(\alpha^2-\alpha-1)f_{1,4}^{(3)} +(6\alpha-4) \left( f_{2,3}^{(3)}-f_{3,3}^{(3)} \right)\,.
\end{align}
Taking $N\to -N$ in the integrand for a giant of dimension $M$ maps $\alpha>0$ to negative values, which corresponds to the dual giant integrand, with the same dimension $M=|\alpha| N$. This is precisely the leading large $N$ realization of the negative rank relation derived in Section~\ref{sec:looprel}.

We remark that, although both here and in \cite{He:2026ios} loop integrands are organized using Lagrangian insertion, the integrands in \cite{He:2026ios} are not obtained from direct Wick contractions. They are determined using an $f$-graph bootstrap with several pieces of OPE and integrated correlator data. Their unified result therefore provides an independent check of our negative rank relation.

\subsection{Partially contracted giant graviton formula}

For perturbative HHLL integrands, it is useful to separate the Wick contractions into two stages. We first contract most of the fields directly between the two heavy operators, leaving the remaining heavy fields to contract with the light operators and interaction vertices. This is known as the partially contracted giant graviton (PCGG) method \cite{Jiang:2019xdz}, which provides a useful technique for computing giant graviton correlators and has been applied, for example, to the two-loop HHLL correlator \cite{Wu:2025ott} and the four-determinant correlator \cite{Vescovi:2021fjf}. Here we provide compact PCGG formulas for giant and dual giant gravitons in the $U(N)$ theory,\footnote{The $SU(N)$ generalization is recorded in Appendix~\ref{app:pcgg}.} and show that, although only partially contracted, the resulting blocks are still related by $N \to -N$.

We begin by introducing a Wick contracting operator
\begin{equation}
  \mathcal C \equiv \frac{\partial}{\partial(\Phi_1)^i{}_j} \frac{\partial}{\partial(\Phi_2)^j{}_i}
\end{equation}
for the heavy fields at points $1$ and $2$. Let $M-\ell$ fields be contracted directly between the heavy operators, leaving $\ell$ fields at each heavy insertion. We call the resulting object a PCGG block,
\begin{equation}
  \mathcal Q_{M,\ell} \equiv \frac{1}{(M-\ell)!} {\mathcal C}^{M-\ell} \bigl[\mathcal D_M(1)\mathcal D_M(2)\bigr]\,.
\end{equation}
Instead of treating each giant operator separately, we consider the generating function $\det(1+t\Phi_1)\det(1+s\Phi_2)$. On each side, we have $M-\ell$ derivatives acting on the determinant,
\begin{equation}
  \frac{\partial^{M-\ell}\det(1+t\Phi_i)}{ \partial(\Phi_i)^{a_1}{}_{b_1}\cdots \partial(\Phi_i)^{a_{M-\ell}}{}_{b_{M-\ell}}}\,.
\end{equation}
These derivatives can be evaluated using the following formula (see Appendix~\ref{app:pcgg} for details)
\begin{equation}
  \frac{\partial^m \det M} {\partial M^{a_1}{}_{b_1}\cdots\partial M^{a_m}{}_{b_m}} = \det M \sum_{\sigma\in S_m} \operatorname{sgn}(\sigma) (M^{-1})^{b_1}{}_{a_{\sigma(1)}}\cdots (M^{-1})^{b_m}{}_{a_{\sigma(m)}}\,.
\end{equation}
Taking the coefficient of $t^M s^M$ to project onto the dimension-$M$ giant operators and contracting all indices, we obtain
\begin{equation}
  \mathcal Q_{M,\ell}={}(M-\ell)! \det(1+t\Phi_1)\det(1+s\Phi_2)\chi_{[1^{M-\ell}]}\!\left((1+t\Phi_1)^{-1}(1+s\Phi_2)^{-1}\right)\Big|_{t^\ell s^\ell}\,, \label{eq:pcgg}
\end{equation}
where taking the coefficient of $t^\ell s^\ell$ extracts the required sub-determinants in the generating function. This PCGG block has $\ell$ remaining fields of $\Phi_1$ and $\Phi_2$ to be contracted with the light operators. In particular, we have the expected values
\begin{equation}
  \mathcal Q_{M,M}=\mathcal D_M(1) \mathcal D_M(2)\,, \qquad \mathcal Q_{M,0}= \frac{N!}{(N-M)!}\,. \label{eq:pcgglim}
\end{equation}
Very similarly, we can derive the PCGG formula for dual giant gravitons. The result
\begin{equation}
  \widetilde{\mathcal Q}_{M,\ell}={}(M-\ell)!\,\det(1-t\Phi_1)^{-1}\det(1-s\Phi_2)^{-1}\chi_{[M-\ell]}\!\left((1-t\Phi_1)^{-1}(1-s\Phi_2)^{-1}\right)\Big|_{t^\ell s^\ell}\,
\end{equation}
is exactly the same as applying $\omega$ on the PCGG blocks for giants
\begin{equation}
  \omega[\mathcal Q_{M,\ell}]=(-1)^{M-\ell}\widetilde{\mathcal Q}_{M,\ell}\,.
\end{equation}

\section{Towards finite and strong couplings}
\label{sec:finite}

We have established the negative rank relation for exact free correlators and for separated point weak coupling integrands at every loop order. In this section we discuss two observations beyond this perturbative setting: a formal analytic continuation between the giant and dual giant geometries at the level of the induced metric, and a finite coupling planar relation for an integrated correlator of \cite{Brown:2025huy}.

\subsection{Giant and dual giant D3 brane geometries}

Following the notation of \cite{Brown:2025huy}, the metric of the $AdS_5\times S^5$ background is
\begin{equation}
  ds^2 = ds_{AdS_5}^2 + ds_{S^5}^2\,,
\end{equation}
with the metrics of $AdS_5$ and $S^5$ given by
\begin{equation}
  \begin{aligned}
  ds_{AdS_5}^2 &= L^2\left( d\rho^2 -\cosh^2\rho\,dt^2 +\sinh^2\rho\,d\widetilde\Omega_3^2 \right)\,, \\
  ds_{S^5}^2 &= L^2\left( d\theta^2+\sin^2\theta\,d\phi^2 +\cos^2\theta\,d\Omega_3^2 \right)\,,
  \end{aligned}
\end{equation}
where $d\Omega_3^2$ and $d\widetilde\Omega_3^2$ are unit three sphere metrics in the $S^5$ and $AdS_5$ respectively. Here we use two positive parameters $\alpha$ and $\beta$ for the sphere giant and dual giant, rather than the signed $\alpha$ in Section~\ref{sec:hhl}. Their embeddings are
\begin{equation}
  \begin{aligned}
  \text{sphere giant:}\qquad &\rho=0\,, &\theta=\theta_0\,,\quad &\phi=t\,,\quad &\alpha=\cos^2\theta_0=\frac{M}{N}\,,\quad &0<\alpha\leq1\,, \\
  \text{dual giant:}\qquad &\theta=\frac{\pi}{2}\,, &\rho=\rho_0\,,\quad &\phi=t\,,\quad &\beta=\sinh^2\rho_0=\frac{M}{N}\,,\quad &\beta>0\,.
  \end{aligned}
\end{equation}
The sphere giant wraps the $S^3$ parametrized by $\Omega_3$ inside $S^5$, while the dual giant instead wraps the $S^3$ parametrized by $\widetilde\Omega_3$ inside $AdS_5$. On the two embeddings, substituting the motion $d\phi=dt$ gives 
\begin{equation}
  \begin{aligned}
  -\cosh^2(0)\,dt^2+\sin^2\theta_0\,d\phi^2 &= -\cos^2\theta_0\,dt^2=-\alpha\,dt^2\,, \\
  -\cosh^2\rho_0\,dt^2+\sin^2\left(\frac{\pi}{2}\right)d\phi^2 &=-\sinh^2\rho_0\,dt^2=-\beta\,dt^2\,.
  \end{aligned}
\end{equation}
Together with the corresponding wrapped sphere components, these pullbacks give the induced metrics on the two D3 branes
\begin{equation}
  ds_{\rm giant}^2 = L^2\alpha\left(-dt^2+d\Omega_3^2\right)\,, \qquad ds_{\rm dual}^2 = L^2\beta\left(-dt^2+d\widetilde\Omega_3^2\right)\,,
\end{equation}
with worldvolume radii $L\sqrt{\alpha}$ and $L\sqrt{\beta}$ respectively. At the level of the metric, we can connect these two solutions by identifying $t$ and $\phi$, and performing the formal Wick rotation
\begin{equation}
  \theta\leftrightarrow\frac{\pi}{2}+i\rho\,,\qquad L^2 \leftrightarrow -L^2\,.
\end{equation}
This gives
\begin{equation}
  \qquad\sin\theta\leftrightarrow\cosh\rho\,, \qquad \cos\theta\leftrightarrow-i\sinh\rho\,, \qquad \alpha\leftrightarrow-\beta\,,
\end{equation}
and switches the roles of $AdS$ and $S$, including the spheres $d\Omega_3^2$ and $d\widetilde\Omega_3^2$ inside
\begin{equation}
  \left.ds_{S^5}^2\right|_{\theta\to\frac{\pi}{2}+i\rho,\ L^2\to -L^2} = -L^2\left( -d\rho^2+\cosh^2\rho\,dt^2 -\sinh^2\rho\,d\widetilde\Omega_3^2 \right) = ds_{AdS_5}^2\,.
\end{equation}
At the level of the induced metric, the giant graviton wrapping inside $S^5$ becomes a dual giant graviton wrapping in $AdS_5$, and vice versa. It would be interesting to understand whether this formal continuation is related to the LLM particle-hole picture \cite{Lin:2004nb}, as discussed in \cite{Berenstein:2020jen,Brown:2025huy,Anempodistov:2025maj}.

It should be noted that this observation concerns only the metric and its pullback to the D3 brane worldvolume. We do not establish an analytic continuation of the full D3 brane action, which would also require treating the RR four-form potential and the Wess-Zumino coupling.

\subsection{Integrated correlator}

Further evidence for the negative rank relation comes from the integrated HHLL correlators studied in \cite{Brown:2025huy}.\footnote{See also \cite{Dorigoni:2022zcr} for related observations in integrated correlators of light operators, with gauge group $SU(N)$ and $SO(N)/Sp(N)$. } They considered two identical sphere giants or two identical dual giants of charge $M$, together with two light operators in the stress tensor multiplet. At leading order in large $N$, the two integrated correlators are functions $\mathcal I_S(\lambda;\alpha)$ and $\mathcal I_{AdS}(\lambda;\beta)$, where we recall that
\begin{equation}
  \alpha=\frac{M}{N}\,,\quad 0<\alpha\leq 1\,,\quad \beta=\frac{M}{N}\,,\quad \beta>0\,,
\end{equation}
for sphere giants and dual giants respectively. The finite coupling results of \cite{Brown:2025huy} obey the relation:\footnote{The expressions below are the $g=0$ terms in the genus expansion $\mathcal I_{\rm full} = \sum_g N^{1-g}\mathcal I^{(g)}$ with superscript omitted. This explains the minus sign in the relation since the full object should be $N\mathcal I_{S/AdS}(\lambda;\alpha)$. }
\begin{equation}
  \mathcal I_S(\lambda;\alpha)=-\mathcal I_{AdS}(\lambda;-\alpha)\,,\quad \alpha<0\,. \label{eq:intrel}
\end{equation}
Here the sphere giant expression on the left-hand side is analytically continued outside its physical domain. This is precisely the continuation expected from the negative rank relation, since $N\to -N$ at fixed $\lambda$ and $M$ reverses the sign of $\alpha=M/N$. The above relation therefore provides evidence for an extension of the negative rank relation to finite coupling.

Note that the condition $\alpha<0$ is not essential, but is imposed just to avoid the need to specify branch choices explicitly. As a function of $\alpha$, the integrated correlator $\mathcal I_S(\lambda;\alpha)$ develops logarithmic singularities at the points (see Appendix~\ref{app:int})
\begin{equation}
  \alpha_n^\pm=\frac{4\pi^2n^2}{\lambda}\pm\frac{4\pi i n}{\sqrt{\lambda}}\,,\quad n=1,2,\ldots\,.
\end{equation}
To equate both sides in \eqref{eq:intrel} as analytic functions over the complex $\alpha$ plane, one must therefore specify compatible branch choices for the singularities of $\mathcal I_S$ and $\mathcal I_{AdS}$.

A subtle feature appears at strong coupling. Although \eqref{eq:intrel} holds for the finite coupling planar functions, the strong coupling expansions in the physical sphere and dual giant regions are not related term by term by $\alpha\to-\beta$. In particular, their leading supergravity contributions are \cite{Brown:2025huy}
\begin{equation}
  \mathcal I_S(\lambda;\alpha) \sim 2\alpha\,,\qquad \mathcal I_{AdS}(\lambda;\beta) \sim 2\log(1+\beta)\,,\quad \lambda \gg 1\,,
\end{equation}
respectively. Moreover, the nonperturbative contributions for the physical giant $0<\alpha\leq 1$ and dual giant $\beta>0$ include terms of the following form
\begin{equation}
  \exp\left[-n\sqrt{\lambda}(1-\sqrt{1-\alpha})\right]\,,\qquad \exp\left[-n\sqrt{\lambda}(\sqrt{1+\beta}-1)\right]\,,\quad n=1,2,\ldots\,.
\end{equation}
Under the naive continuation $\alpha\to-\beta$, an exponentially suppressed term is mapped to an exponentially growing one. The physical giant and dual giant expansions therefore belong to different asymptotic sectors, as expected in the presence of a Stokes phenomenon. Thus the analytic continuation of the finite coupling functions is not captured by a termwise continuation of their strong coupling expansions, and we need to be very careful about the negative rank relation in the strong coupling expansions.

We emphasize that, since \eqref{eq:intrel} concerns a particular integrated correlator, it does not by itself establish a negative rank relation for the full finite coupling correlator.

\acknowledgments
We thank Yunfeng Jiang, Chang Liu, Yichao Tang and Congkao Wen for helpful comments on the draft, and Yunfeng Jiang for collaboration on related projects. This work used OpenAI Codex to assist with literature searches, exploratory calculations and language editing. All AI outputs were independently cross-checked and verified. The work of X.Z. is supported by the NSFC Grant No. 12275273 and funds from the Chinese Academy of Sciences. Z.H. is supported by the NSFC Grant No. 12547189.

\appendix

\section{Details of PCGG formulas}
\label{app:pcgg}

We provide here some useful details for the PCGG formulas, and generalize them to the $SU(N)$ theory.

\subsection{Derivative formulas}

Repeated derivatives of a determinant can be evaluated recursively from the following three seed formulas
\begin{align}
  \frac{\partial \det M}{\partial M^a{}_b} &=\det M(M^{-1})^b{}_a\,, \\
  \frac{\partial \det M^{-1}}{\partial M^a{}_b} &=-\det M^{-1}(M^{-1})^b{}_a\,, \\
  \frac{\partial (M^{-1})^b{}_a}{\partial M^c{}_d} &=-(M^{-1})^b{}_c(M^{-1})^d{}_a\,. \label{eq:dm}
\end{align}
Each additional derivative can hit either the determinant or an existing factor of $M^{-1}$. Hitting the determinant leaves the matrix indices unchanged, whereas hitting $M^{-1}$ exchanges the indices of two matrices and introduces a minus sign. So, recursively, by Leibniz's rule we need to sum over all the ways of taking derivatives, which generates all possible permutations $\sigma\in S_m$ on the indices, with a $\operatorname{sgn}(\sigma)$ if the minus sign in \eqref{eq:dm} is not canceled. We therefore have
\begin{align}
  \frac{\partial^m \det M} {\partial M^{a_1}{}_{b_1}\cdots\partial M^{a_m}{}_{b_m}} &= \det M \sum_{\sigma\in S_m} \operatorname{sgn}(\sigma) \prod_{r=1}^m (M^{-1})^{b_r}{}_{a_{\sigma(r)}}\,, \label{eq:ddet}\\
  \frac{\partial^m (\det M)^{-1}} {\partial M^{a_1}{}_{b_1}\cdots\partial M^{a_m}{}_{b_m}} &= (-1)^m(\det M)^{-1} \sum_{\sigma\in S_m} \prod_{r=1}^m (M^{-1})^{b_r}{}_{a_{\sigma(r)}}\,. \label{eq:dinvdet}
\end{align}

There is also a useful auxiliary field interpretation of these two derivative formulas. We take the simpler example \eqref{eq:dinvdet}. We can write the inverse determinant as a Gaussian integral,
\begin{equation}
  (\det M)^{-1}\propto \int d^N z\,d^N\bar z\, \exp\!\left(-\bar z_a M^a{}_bz^b\right)\,,
\end{equation}
and a derivative with respect to $M^a{}_b$ inserts $-\bar z_a z^b$. Hence the $m$-th derivative is proportional to a $2m$-point function of fields $z$ and $\bar z$. Wick contraction gives
\begin{equation}
  \left\langle z^{b_1}\cdots z^{b_m} \bar z_{a_1}\cdots\bar z_{a_m} \right\rangle = \sum_{\sigma\in S_m} \prod_{r=1}^m (M^{-1})^{b_r}{}_{a_{\sigma(r)}}\,,
\end{equation}
which immediately reproduces \eqref{eq:dinvdet}. For \eqref{eq:ddet} we can introduce fermion fields $\chi$ and $\bar\chi$, which is exactly what is done in the large $N$ effective theory for giant gravitons in \cite{Jiang:2019xdz}.

\subsection{Recursive trace expansion for the PCGG blocks}

The generating function \eqref{eq:pcgg} gives a compact expression for the general block. For explicit applications, it is useful to expand it in cyclic trace words built from the remaining fields $\Phi_1$ and $\Phi_2$ at the two heavy insertions. In particular, for an efficient implementation, all the terms should be recursively determined. We now give a finite recursion which performs this expansion.

We first rewrite the PCGG block \eqref{eq:pcgg} as:\footnote{Taking $M=N$ this reduces to the PCGG formula in \cite{Jiang:2019xdz} for maximal giant gravitons
\begin{equation}
  \mathcal Q_{N,\ell}=(N-\ell)!\chi_{[1^\ell]}(\Phi_2\Phi_1) = (N-\ell)!(-1)^\ell\sum_{\substack{k_1,\dots,k_\ell\\\sum_s sk_s=\ell}} \prod_{m=1}^\ell \frac{(-\tr \left[\left(\Phi_2\Phi_1)^m\right]\right)^{k_m}}{m^{k_m}k_m!}\,.
\end{equation}
}
\begin{equation}
  \mathcal Q_{M,\ell}={}(M-\ell)!\,\chi_{[1^{N-M+\ell}]}\!\left((1+s\Phi_2)(1+t\Phi_1)\right)\Big|_{t^\ell s^\ell}\,,
\end{equation}
using
\begin{equation}
  \det X \chi_{[1^n]}(X^{-1})=\chi_{[1^{N-n}]}(X)\,.
\end{equation}
Denoting $ H = s\Phi_2 + t\Phi_1 + st\Phi_2\Phi_1$, we can further impose the identity shift formula
\begin{equation}
  \chi_{[1^n]}(1+H) = \sum_{m=0}^n \binom{N-m}{n-m} \chi_{[1^m]}(H)\,,
\end{equation}
and use Newton's identity to turn recursively the Schur polynomials into traces
\begin{equation}
  \chi_{[1^m]}(H) = \sum_{q=1}^m \frac{(-1)^{q-1}}{m}\chi_{[1^{m-q}]}(H)\tr H^q\,.
\end{equation}

We still need to expand $H$ and extract the $t^\ell s^\ell$ coefficient. For convenience we introduce
\begin{align}
  E_{a,b,c} &\equiv \chi_{[1^{a+b+c}]}\, \left(s\Phi_2+t\Phi_1+u\,\Phi_2\Phi_1\right)\Big|_{s^at^bu^c}\,,\\
  W_{a,b,c} &\equiv  \left(s\Phi_2+t\Phi_1+u\,\Phi_2\Phi_1\right)^{a+b+c}\Big|_{s^at^bu^c}\,,
\end{align}
and set $u=st$ later. We can thus write down the recursion relations
\begin{align}
  E_{a,b,c} &= \sum_{\substack{0\leq i,j,k\leq a,b,c\\(i,j,k)\neq(0,0,0)}} \frac{(-1)^{i+j+k-1}}{a+b+c} E_{a-i,b-j,c-k}\, \tr W_{i,j,k}\,, \\
  W_{a,b,c} &= W_{a-1,b,c} \Phi_2 + W_{a,b-1,c} \Phi_1 + W_{a,b,c-1} \Phi_2\Phi_1\,,
\end{align}
with the initial conditions $E_{0,0,0}=1$ and $W_{0,0,0} = \mathbf 1$. Putting all these ingredients together we obtain the following sum
\begin{equation}
  \mathcal Q_{M,\ell} = \sum_{k=0}^\ell (N-M-k+1)_{M-\ell}\, E_{k,k,\ell-k}\,.
\end{equation}
All explicit dependence on $M$ and $N$ is contained in the Pochhammer coefficients. The corresponding PCGG blocks for dual giant gravitons can be thus written as
\begin{equation}
  \widetilde{\mathcal Q}_{M,\ell} = \sum_{k=0}^\ell (N+\ell+k)_{M-\ell}\, \omega[E_{k,k,\ell-k}]\,.
\end{equation}

\subsection{The \texorpdfstring{$SU(N)$}{SU(N)} extension of the PCGG blocks}

The PCGG blocks were constructed using the $U(N)$ propagator. We now implement the traceless projection directly in the partial contraction between the two heavy insertions and express the resulting $SU(N)$ blocks as finite sums of the $U(N)$ ones. This was considered in \cite{Brown:2024tru} for maximal giants at subleading order in the large $N$ expansion. Here we will provide a finite $N$ formula for arbitrary charges.

The color part of the $SU(N)$ propagator differs from the $U(N)$ one by the trace subtraction in \eqref{eq:suprop}. Accordingly, we define the projected Wick contracting operator
\begin{equation}
  \widehat{\mathcal C} \equiv \mathcal C -\frac{1}{N}\left( \tr \frac{\partial}{\partial\Phi_1} \right) \left( \tr \frac{\partial}{\partial\Phi_2} \right)\,,\quad \tr \frac{\partial}{\partial\Phi_i} \equiv \frac{\partial}{\partial(\Phi_i)^a{}_a}\,.
\end{equation}
Indeed,
\begin{equation}
  \widehat{\mathcal C} \left[ (\Phi_1)^a{}_b(\Phi_2)^c{}_d \right] = \delta^a{}_d\delta^c{}_b -\frac{1}{N}\delta^a{}_b\delta^c{}_d\,,
\end{equation}
which reproduces the color structure of the traceless propagator.

The PCGG blocks for the $SU(N)$ gauge group are therefore
\begin{equation}
  \begin{aligned}
  \mathcal Q_{M,\ell}^{SU}(N) &\equiv \frac{1}{(M-\ell)!} \widehat{\mathcal C}^{M-\ell} \bigl[ \mathcal D_M(1)\mathcal D_M(2) \bigr]\,, \\
  \widetilde{\mathcal Q}_{M,\ell}^{SU}(N) &\equiv \frac{1}{(M-\ell)!} \widehat{\mathcal C}^{M-\ell} \bigl[ \widetilde{\mathcal D}_M(1) \widetilde{\mathcal D}_M(2) \bigr]\,.
  \end{aligned}
\end{equation}
As in Section~\ref{sec:su}, the derivatives are first evaluated by treating the matrix entries of $\Phi_i$ as $U(N)$. Only afterward are the remaining fields restricted to the traceless $SU(N)$ fields.

We may expand
\begin{equation}
  \begin{aligned}
  \widehat{\mathcal C}^{M-\ell} ={}& \sum_{r=0}^{M-\ell} \binom{M-\ell}{r} \frac{(-1)^r}{N^r} \mathcal C^{M-\ell-r} \left( \tr \frac{\partial}{\partial\Phi_1} \right)^r \left( \tr \frac{\partial}{\partial\Phi_2} \right)^r\,.
  \end{aligned}
  \label{eq:subinom}
\end{equation}
The trace derivatives lower the heavy charge. Acting once on the giant and dual giant generating functions gives
\begin{equation}
  \begin{aligned}
  \tr \frac{\partial}{\partial\Phi_i}\det(1+t\Phi_i) &=t\det(1+t\Phi_i) \tr (1+t\Phi_i)^{-1}=\sum_{k\geq1}(N-k+1)\mathcal D_{k-1}(i)t^k\,, \\
  \tr \frac{\partial}{\partial\Phi_i} \det(1-t\Phi_i)^{-1} &= t\det(1-t\Phi_i)^{-1} \tr (1-t\Phi_i)^{-1}=\sum_{k\geq1} (N+k-1)\widetilde{\mathcal D}_{k-1}(i)t^k\,.
  \end{aligned}
\end{equation}
Comparing coefficients and iterating yields
\begin{equation}
  \begin{aligned}
  \left( \tr \frac{\partial}{\partial\Phi_i} \right)^r \mathcal D_M(i) &=(N-M+1)_r\mathcal D_{M-r}(i)\,, \\
  \left(\tr \frac{\partial}{\partial\Phi_i} \right)^r   \widetilde{\mathcal D}_M(i) &=(N+M-r)_r\widetilde{\mathcal D}_{M-r}(i)\,.
  \end{aligned}
  \label{eq:lower}
\end{equation}
These reduce the PCGG blocks to lower charges $M-r$. Substituting \eqref{eq:lower} into \eqref{eq:subinom} therefore gives
\begin{equation}
  \begin{aligned}
  \mathcal Q_{M,\ell}^{SU}(N) &=\sum_{r=0}^{M-\ell} \frac{(-1)^r}{r!N^r} (N-M+1)_r^2 \mathcal Q_{M-r,\ell}\,, \\
  \widetilde{\mathcal Q}_{M,\ell}^{SU}(N) &=\sum_{r=0}^{M-\ell} \frac{(-1)^r}{r!N^r}  (N+M-r)_r^2  \widetilde{\mathcal Q}_{M-r,\ell}\,.
  \end{aligned}
  \label{eq:supcgg}
\end{equation}
Note that \eqref{eq:supcgg} accounts only for the direct contractions between the two heavy operators. Any later contraction of the remaining fields with light operators or interaction vertices must still be performed using the $SU(N)$ propagator.

As a simple consistency check, we set $\ell=0$, so that all heavy fields are contracted between the two insertions. Using \eqref{eq:pcgglim} in \eqref{eq:supcgg} gives
\begin{equation}
  \begin{aligned}
  \mathcal Q_{M,0}^{SU}(N) &= \sum_{r=0}^M \frac{(-1)^r}{r!N^r}(N-M+1)_r^2   \frac{N!}{(N-M+r)!}\,, \\
  \widetilde{\mathcal Q}_{M,0}^{SU}(N) &= \sum_{r=0}^M \frac{(-1)^r}{r!N^r}(N+M-r)_r^2 (N)_{M-r}\,.
  \end{aligned}
\end{equation}
These are precisely the $SU(N)$ two-point functions for the giant and dual giant gravitons \cite{deMelloKoch:2004crq}. One can also check that
\begin{equation}
  \omega[\mathcal Q_{M,\ell}^{SU}] =(-1)^{M-\ell}\widetilde{\mathcal Q}_{M,\ell}^{SU}\,.
\end{equation}

\section{Singularities in the integrated HHLL correlator} \label{app:int}

Following \cite{Brown:2025huy}, the integrated correlator for the giant graviton, $\mathcal I_S(\lambda;\alpha)$, is defined by the analytic continuation of
\begin{align}
  \mathcal I_S(\lambda;\alpha) = \mathcal I_{S; 0}(\lambda) + \mathcal I_{S; 1}(\lambda;\alpha) \,,
\end{align}
where
\begin{align}
  \mathcal I_{S; 0}(\lambda) &= 4 \sum_{\ell=1}^\infty (-1)^{\ell+1} \,\zeta(2\ell+1)\Big(\frac{\lambda}{16\pi^2}\Big)^\ell \binom{2\ell{+}1}{\ell}^2 \,, \\
  \mathcal I_{S; 1}(\lambda;\alpha) &= 4 \sum_{\ell=1}^\infty (-1)^\ell \,\zeta(2\ell+1)\Big(\frac{\lambda}{16\pi^2}\Big)^\ell  \alpha^{2(\ell+1)} \binom{2\ell {+} 1}{\ell}  \, _2 F_1 \big(\ell+1,\ell+2;1;1-\alpha\big) \,.
\end{align}
At fixed $\lambda$, the singularities in $\alpha$ are controlled by the large $\ell$ tail of the infinite sum. Since $\mathcal I_{S;0}$ is independent of $\alpha$, only $\mathcal I_{S;1}$ contributes to these singularities. We therefore follow the analysis in \cite{Brown:2025huy} and use the following large $\ell$ asymptotics\footnote{The equation (4.16) of \cite{Brown:2025huy} considered only the $\sigma=+$ branch, which is leading at $0 \leq \alpha < 1$.}
\begin{align}
  \binom{2\ell+1}{\ell} &\sim \frac{2^{2\ell +1}}{\sqrt{\pi \ell}} \,, \qquad \ell \gg 1\,,\\
  \alpha^{2(\ell+1)}\, _2 F_1 \big(\ell+1,\ell+2;1;1-\alpha\big) &\sim   \sum_{\sigma=\pm} \frac{\left(1+ \sigma\sqrt{1-\alpha }\right)^{2 (\ell+1)} }{2\,  \sqrt{\pi \, \ell}\, \sqrt{\sigma \sqrt{1-\alpha} } }\,, \qquad \ell\gg 1\,.
\end{align}
By using
\begin{equation}
  \zeta(2\ell+1) = \sum_{n=1}^\infty\frac{1}{n^{2\ell+1}}\,,
\end{equation}
we may consider each fixed $n$ separately. Up to overall coefficients, the large $\ell$ tail of the second term then behaves as
\begin{align}
  \sum_{\ell\gg 1}\frac{(-1)^\ell}{\ell} \frac{1}{n^{2\ell}}\left( \frac{\lambda}{16\pi^2} \right)^\ell 2^{2\ell} \left(1 + \sigma \sqrt{1-\alpha }\right)^{2 \ell}  \sim  \log \left[1+\left(\frac{\sqrt{\lambda} \left(1+ \sigma \sqrt{1-\alpha}\right)}{2\pi n}\right)^2\right]\,.
\end{align}
The logarithm therefore develops branch points when
\begin{equation}
  1+ \left( \frac{ \sqrt{\lambda}\left(1+\sigma\sqrt{1-\alpha}\right) }{2\pi n} \right)^2 =0\,.
\end{equation}
Combining both $\sigma$ branches gives the infinite set of branch points 
\begin{equation}
  \alpha_n^\pm = \frac{4\pi^2n^2}{\lambda} \pm \frac{4\pi i n}{\sqrt{\lambda}}\,, \qquad n=1,2,\ldots\,,
\end{equation}
in the complex $\alpha$ plane.

\bibliography{refs}
\bibliographystyle{utphys}
\end{document}